\documentstyle[12pt,titlepage,a4,epsf]{article}

\renewcommand{\thesubsection}{\arabic{subsection}~$\!\!\!\!$}
\newcounter{sub}

\begin{document}
\begin{titlepage}
\begin{flushright}
  {\tt hep-th/9612237}\\
  $\begin{array}{lr}
  \mbox{\small UT-KOMABA/96-28} \!&\!\! \mbox{\small NUP-A-96-14}\\ 
  \mbox{\small NBI-HE-96-65}    \!&\!\! \mbox{\small TIT/HEP-351}
  \end{array}{\!\!\!\!\;}$\\
  December 1996\\
\end{flushright}

\begin{center}
{\large\bf
Renormalization group approach\\
to multiple-arc random matrix models
}
\vspace*{\fill}

{\sc Saburo Higuchi}
\footnote{{\tt e-mail: hig@rice.c.u-tokyo.ac.jp}}\\
{\it Department of Pure and Applied Sciences, 
University of Tokyo \\
Komaba, Meguro, Tokyo 153, Japan} 
\bigskip

{\sc Chigak Itoi}
\footnote{\tt e-mail: itoi@phys.cst.nihon-u.ac.jp} \\
{\it Department of Physics,
College of Science and Technology, 
Nihon University \\
Kanda Surugadai, Chiyoda, Tokyo 101, Japan}
\bigskip

{\sc Shinsuke M. Nishigaki}
\footnote{\tt e-mail: shinsuke@alf.nbi.dk}\\
{\it Niels Bohr Institute\\
Blegdamsvej 17, DK-2100 Copenhagen \O, Denmark
}
\bigskip

{\sc Norisuke Sakai}
\footnote{\tt e-mail: nsakai@th.phys.titech.ac.jp} \\
{\it Department of Physics,
Tokyo Institute of Technology \\
Oh-okayama, Meguro, Tokyo 152, Japan}

\vspace*{\fill}
{\bf Abstract}
\end{center}
  We study critical and 
  universal behaviors of unitary invariant
  non-gaussian random matrix ensembles within 
  the framework of the large-$N$ renormalization group.
  For a simple double-well model
  we find an unstable fixed point and
  a stable inverse-gaussian fixed point.
  The former is identified as the critical point of
  single/double-arc phase transition
  with a discontinuity of
  the third derivative of the free energy.
  The latter signifies a novel universality 
  of large-$N$ correlators other than the
  usual single arc type.
  This phase structure is consistent with the
  universality classification of
  two-level correlators for multiple-arc models
  by Ambj{\o}rn and Akemann.
  We also establish the stability of the gaussian fixed
  point in the multi-coupling model.\\[4mm]
\noindent
PACS: 11.15.Pg, 05.45.+b
\vfill
\end{titlepage}
\setcounter{footnote}{0}

\subsection{Introduction}
The universal behaviors  
in random matrix theories are key ingredients in their
level-statistical application, and have been discussed extensively.
It is known that there are several universal quantities
which do not explicitly depend on the probability distribution of 
a matrix ensemble. 
Because of the universality
we may calculate such quantities from a simple gaussian ensemble 
which shares the same symmetries as 
the physical system in concern.
Recently, Ambj{\o}rn and Akemann pointed out 
a universality classification of matrix ensembles 
with respect to the two-level correlators \cite{AmAk:newuniversal}.
The large-$N$ (or ``smoothed'' \cite{BrZe:universal})
connected two-level correlator 
depends solely on the support configuration of the level density 
$\rho(\lambda)$.
For example, this correlator in a
matrix ensemble with a single-arc level density 
is identical to that calculated in the gaussian ensemble.

In this letter, we study the phase structure of a non-gaussian 
matrix ensemble with a multiple-arc level density
in terms of the large-$N$ renormalization group 
\cite{BrZi:RGMat}.
The general algorithm of the 
large-$N$ renormalization group 
enables us to find out the exact location of 
the critical point as a
fixed point and to specify the order of 
the phase transition exactly
\cite{HiItNiSa:Matrix}. 
Moreover, its renormalization group flow clarifies 
the global phase structure
by classifying the parameter space into
phases separated by critical surfaces emanating
from unstable fixed points
\cite{HiItNiSa:oneandtwo}.
Along this line,
a conjectural account for the branched-polymeric behavior of
$c>1$ bosonic strings
was recently provided \cite{David:c>1}.

We first concern ourselves on the merging phase transition where
two energy bands coalesce 
into a single band.
This merging transition has recently attracted attention, 
for instance,
in the application of random matrix theory
to the level statistics of QCD Dirac operator 
\cite{ShVe:RMTsumrule};
there the level density around $\lambda=0$ 
is identified with the chiral condensate of quarks,
and thus its vanishing 
is related to the chiral symmetry
restoration at a finite temperature \cite{JaSeVe:zerovertuality}.
More specifically, in order to study the phase structure
involving different arc configurations,
we investigate the simplest non-gaussian model with a
U($N$) and ${\bf Z}_2$-invariant distribution 
$
P(\phi) \propto \exp 
\left[-N \,{\rm tr}\, (-\frac12\phi^2+ \frac{g}{4}\phi^4) \right]
$, 
which can take either single or double-arc 
level density depending on the parameter $g$.
It is known that
at a merging point $g= 1/4$,
the third derivative of the free energy 
has a discontinuity due to the switchover of the functional form
of $\rho(\lambda)$
\cite{Shimamune:phasesep}. 
We shall correctly identify this critical point 
as an unstable fixed point of the renormalization group,
from which the renormalization group flow is emitted. 
As a straightforward exercise
we derive the explicit form of the free energy 
by integrating the renormalization group equation. 

Next we shall discuss the off-critical,
universal behavior of the model.
It is argued to be
determined by the nature around the stable fixed point
which is a final destination of the renormalization group flow
\cite{BrZe:universal}.
To proceed we need to define the approximate $\beta$-function
with required properties. 
The phase diagram described by such a $\beta$-function
is in accord with
the above mentioned universality classification
by the large-$N$ two-level correlator.  
Finally we establish the stability of the gaussian point 
by calculating all the associated scaling exponents
in multi-coupling models.

\subsection{Renormalization group approach}
We consider an ensemble of random $N\times N $ 
hermitian matrices $\phi$ with a weight
\begin{equation}
  P_N(\phi)  =  Z_N(g_j)^{-1} \exp[-N \,{\rm tr}\, V(\phi)], 
  \ \ \ 
  V(\phi)   =  \sum_{k=2}^m \frac{g_k}{k}\phi^k.
  \label{potential}
\end{equation}
Hereafter the coupling constant $g_2$ is fixed 
either to $-1$ (in sect.3, 4, 5) or 
to $+1$ (in sect.6),
and $g_2$-dependences 
will not be indicated explicitly.
The expectation value
with respect to the weight is denoted by 
$\langle \cdots \rangle$.
The partition function $Z_N(g_j)$ is defined so that
$\langle 1 \rangle = 1$. 
It is written as an integral over eigenvalues
\begin{equation}
Z_{N}(g_j) 
 = c_{N}\int \prod_{\ell=1}^{N} d\lambda_\ell
\prod_{1 \leq i < j \leq N} \left| \lambda_i-\lambda_j \right|^2
\exp\left[-N\sum_{k=1}^N V(\lambda_k)\right]
\label{partitition-eigenvalue}
\end{equation}
after an integration over angular part of $\phi$,
which leaves us the volume $c_N$ of U$(N)$. 
The free energy is defined as
\begin{equation}
  F(N,g_j)=-\frac{1}{N^2} 
\log\left[\frac{Z_N(g_j)}{Z_N(g_j=0)|_{g_2=1}}\right].
\label{free-energy}
\end{equation}

We brief\/ly recall the renormalization group approach to random
matrix models developed in 
refs.\cite{BrZi:RGMat,HiItNiSa:Matrix,HiItNiSa:oneandtwo}.
In the spirit of the renormalization group,
we integrate a part of the degrees 
of freedom, namely the $(N+1)$-th
eigenvalue $\lambda_{N+1}\equiv \lambda$, 
in $Z_{N+1}(g_j)$ in order to relate it to $Z_N(g_j+\delta g_j)$,
\begin{eqnarray}
\!&\! 
\! &\!
  Z_{N+1}(g_j) 
=
 \frac{c_{N+1}}{c_N} \int d\phi 
\,{\rm e}^{-(N+1)\,{\rm tr}\, V(\phi)} 
\int d\lambda\, {\rm e}^{-(N+1) V(\lambda) 
+ 2\,{\rm tr}\, \log|\lambda -\phi|}\nonumber \\
\!&\! 
= 
\!&\! \frac{c_{N+1}}{c_N} Z_N(g_j)
\exp \left[
-\langle \,{\rm tr}\, V(\phi) \rangle 
- NV(\langle \lambda_{\rm s} \rangle)
 + 2 \langle \,{\rm tr}\, \log | 
\langle \lambda_{\rm s} \rangle - \phi| \rangle
\right].
\label{Zn+1}
\end{eqnarray}
In the second line,
we have evaluated
the $\lambda$-integration
by the large-$N$ saddle point,
and used the factorization property of the 
correlation functions.
It leads to 
the saddle point equation for 
the expectation value
$\langle\lambda_{\rm s}(g_j,\phi)\rangle$:
\begin{equation}
  V'(\langle\lambda_{\rm s}\rangle) 
= 2 \left\langle 
\frac1N \,{\rm tr}\, \frac{1}{\langle\lambda_{\rm s}\rangle-\phi}
\right\rangle.\label{spe}
\end{equation}
By the use of the loop (Schwinger-Dyson) equation, eq.(\ref{spe})
is shown to be equivalent to the condition
\begin{equation}
\rho(\langle \lambda_{\rm s} \rangle)=0,
\ \ \ \ \ 
\left( \rho(
\lambda):=
-\frac1\pi\, {\rm Im}
\left\langle 
\frac1N {\,{\rm tr}\,}\frac{1}{\lambda+i \epsilon-\phi}
\right\rangle 
\right)
\end{equation}
which means that the $(N+1)$-th eigenvalue
settles down to one of the edges of the arcs
of $\rho(\lambda)$.
Expanding eq.(\ref{Zn+1}) into a series in $1/N$,
we have the renormalization group equation obeyed by the free
energy:
\begin{eqnarray}
\lefteqn{\left[ N \frac{\partial}{\partial N} + 2 \right] F(N,g_j)} 
\nonumber \\  
& =&  -\frac32 + 
\left\langle\frac1N \,{\rm tr}\, V(\phi) \right\rangle
 + V(\langle\lambda_{\rm s}\rangle) 
 - 2 \left\langle \frac1N \,{\rm tr}\, \log | 
\langle \lambda_{\rm s} \rangle-\phi|
     \right\rangle + O\left(\frac1N\right) \nonumber \\
& =: &  G\left(g_j, \frac{\partial F}{\partial g_j}
\right).
\label{nrge}
\end{eqnarray}
In the right hand side 
the coupling constants $g_3,\ldots, g_m$ and 
the one-point functions 
$a_j:=\frac{\partial F}{\partial g_j} 
= \frac1j\langle \frac1N \,{\rm tr}\, \phi^j \rangle$ 
$(3\leq j \leq m)$ 
are regarded as independent variables.
Other one-point functions 
appearing in the right hand side 
are understood as  functions of $g_j$ and $a_j$ $(3\le j\le m)$ 
via the 
Schwinger-Dyson equations.
In the same way, $\langle \lambda_{\rm s} \rangle$ is also
regarded as a function of   
$g_j$, $a_j$ $(3 \le j\le m)$.
The explicit form of $G$ is given by 
\begin{eqnarray}
\lefteqn{  G(g_j,a_j)
= -1 + \sum_{j=3}^m\left(1-\frac{j}{2}\right)g_ja_j 
  + V(\langle \lambda_{\rm s} \rangle)} \label{g-function} \\
  & - & 2 \int^{\langle\lambda_{\rm s}\rangle}_{\pm\infty} dz
\left(  \left\langle 
\frac1N \,{\rm tr}\, \frac{1}{z-\phi}\right\rangle -\frac1z\right)
  - 2 \log \langle \lambda_{\rm s} \rangle.
\nonumber
\end{eqnarray}

Let us turn to the process of extracting singular behavior of 
$F$ from the renormalization group equation (\ref{nrge}).
For simplicity, 
we consider the case with a single coupling constant $g$.
The extension to multi-coupling constant case is 
straightforward \cite{HiItNiSa:oneandtwo}.
We assume that 
the free energy in the large-$N$ limit has a 
fractional power singularity 
\begin{equation}
  F(g,N) = \sum_{k=0}^\infty a_k(g-g_*)^k 
        + \sum_{k=0}^\infty b_k(g-g_*)^{k+\gamma} + O(N^{-2})
  \ \ (\gamma \in {\bf R}_+\setminus{\bf Z}, b_0\ne0).
       \label{expandfractional}
\end{equation}
The critical point $g_*$, critical exponent $\gamma$
and coefficients $a_1, a_2$
are determined by solving the following closed set of equations
\renewcommand{\theequation}{\arabic{equation}\alph{sub}}
\begin{eqnarray}
  0 & = & G_{,a}(g_*,a_1) \ ,
\label{fixedpointcond} \\
\addtocounter{equation}{-1}
\addtocounter{sub}{1}
  \frac{2}{\gamma} &=& G_{,g,a}(g_*,a_1) + 2 a_2\,G_{,a,a}(g_*,a_1)\ ,
\label{exponentcond} \\
\addtocounter{equation}{-1}
\addtocounter{sub}{1}
  2 a_1 & = & G_{,g}(g_*,a_1) \ ,
\label{regular1} \\
\addtocounter{equation}{-1}
\addtocounter{sub}{1}
  2 a_2 & = & \frac12 G_{,g,g}(g_*,a_1) + 2a_2\,G_{,g,a}(g_*,a_1) 
                 + 2 (a_2)^2 G_{,a,a}(g_*,a_1)\ , 
\label{regular2} 
\end{eqnarray}
where ${}_{,g}$ and ${}_{,a}$ denote partial derivatives.
These condition can easily be established by substituting
(\ref{expandfractional}) into the
large-$N$ renormalization group equation
\renewcommand{\theequation}{\arabic{equation}}
\begin{equation}
  2 F(N,g) = G\left(g, \frac{dF}{dg} \right).
\end{equation}
By comparing coefficients of $(g-g_*)^k$ and 
$(g-g_*)^{k+\gamma}$ in the both sides 
we obtain a family of
equations for $a_j, b_j, g_*$, and $\gamma$.
The coefficients of $(g-g_*)^{k+\gamma}$ contain both $a_j$ and $b_j$, 
whereas those of $(g-g_*)^{k}$ do only $a_j$ (thus they can
be solved for $a_j$).
Since these equations are homogeneous in $b_0$
the overall normalization of the singular part is left
undetermined.

The above procedure can as well be used to detect the discontinuity 
of a derivative of the free energy.
We now assume that the $\gamma$-th derivative of the free energy is
discontinuous: 
\begin{eqnarray}
\!\!\!&\!\!\!&\!\!\!  
F(g,N) = \sum_{k=0}^\infty a_k(g-g_*)^k 
+ \sum_{k=0}^\infty b_k (g-g_*)^{k+\gamma}\mbox{sgn}[g-g_*] +
O(N^{-2})
\label{expanddiscontinuity}
\\
\!\!\!&\!\!\!&\!\!\!  
\hspace{256pt}
   (\gamma \in {\bf Z}_+, b_0\ne0).
\nonumber
\end{eqnarray}
By substituting (\ref{expanddiscontinuity}) into the
large-$N$ renormalization group equation
and comparing coefficients of $(g-g_*)^k$ and 
$(g-g_*)^{k} \mbox{sgn}[g-g_*]$,
we can easily show that eqs.(10) hold\footnote{
So far we have implicitly assumed that  
the powers of the singular parts do not contribute to the
regular parts. 
It is true for $ \gamma \not \in {\bf Q}$ or
$\gamma=p/q$ with $p-q\ge3$ and $p,q$ are co-prime
(fractional power singularity) 
and for $\gamma \ge3$
(discontinuity)
.}.

\subsection{Merging transition fixed point}
The simplest model which undergoes the merging transition is the one
with quartic potential with negative quadratic term
\cite{Shimamune:phasesep}
\begin{equation}
  V(\phi) = -\frac12 \phi^2 + \frac{g}{4} \phi^4.
  \label{shimamune}
\end{equation}
Though a parameter $g_3$ is missing compared to (\ref{potential}),
we can set $g_{\rm odd} = a_{\rm odd}=0$ consistently because of
the ${\bf Z}_2$ symmetry\footnote{Here we have excluded 
the possibility of spontaneous
breaking of the ${\bf Z}_2$ symmetry \cite{Shimamune:phasesep}.
The linearized RG flow for the broken ${\bf Z}_2$ phase is readily
found in Fig.3 of ref.\cite{HiItNiSa:oneandtwo}
by restricting it onto the curve $2 g_3^2-9g_4=0$.}.

We present the explicit form of the $G$-function:
\begin{eqnarray}
&&G(g,a)  =  
-1 - ga - \frac12 \langle \lambda_{\rm s} \rangle^2 + 
\frac{g}{4}\langle \lambda_{\rm s} \rangle^4  
-2\log \langle \lambda_{\rm s} \rangle 
\label{G}
\\
&& -  \int^{\langle \lambda_{\rm s} \rangle}_{\pm \infty} dz \,
\left(-z+gz^3- \sqrt{(-z+gz^3)^2 - 4 (-1+gz^2-g+4g^2a)}-\frac2z \right)
\nonumber
\end{eqnarray}
which is supplemented by the saddle point equation:
\begin{equation}
  (-\langle \lambda_{\rm s} \rangle + g\langle \lambda_{\rm s}
\rangle^3)^2 - 
4 (-1 + g\langle \lambda_{\rm s} \rangle ^2 -g + 4 g^2 a)=0
\label{saddle}
\end{equation}
determining $\langle \lambda_{\rm s} \rangle$ as a function of $g$
and $a$.

One can show analytically that
$g_*=1/4, \gamma = 3, a_1=5, a_2=-18$ solves 
eqs.(10).
The integer value
 of the $\gamma$ exponent suggests
that around $g_*=1/4$ 
the free energy is of the latter form (\ref{expanddiscontinuity}) 
and has a discontinuity in its third derivative..
This is in accord with the  large-$N$ solution found in 
\cite{Shimamune:phasesep};
at $g \searrow 1/4$, the double-well potential becomes well separated
so that the level density splits into two disconnected arcs.
We stress that our renormalization group equation can actually
detect not only the fractional power behavior of $F(g)$
but also its discontinuity
due to the switchover of the functional form of $\rho(\lambda)$.
 
The readers should notice that,
in the previous procedure
we have not made use of any kinds of
multiple-arc ansatz, which is
used in obtaining large-$N$ solutions
\cite{BrItPaZu:planar}.
If we were interested only in a particular arc configuration, 
we could assume a factorized form of the discriminant of
the square root in the $G$-function (\ref{G}). 
However our method,
which uses the discriminant containing a variable $a$
as it is, can represent any possible arc configurations.
Thus by exploring all solutions to
eqs.(10)
in the two-parameter space $(g,a)$,
we can find solutions with arbitrary arc configurations
respecting the ${\bf Z}_2$ symmetry.

\subsection{Free energy}
To obtain the free energy, one usually
reads off the level density
$\rho(\lambda)$ from the 
expectation value of the resolvent and then takes the average of 
$V(\lambda)-{\cal P}
\log|\lambda-\nolinebreak[1] \lambda'|$ with respect to it
\cite{BrItPaZu:planar}.
Here we show that the explicit form of the free energy can
as well be obtained 
by integrating the renormalization group equation (\ref{nrge}). 
Our calculation avoids computing cumbersome
principal-part integration.

For this practical use,
it suffices to choose a particular arc configuration from the outset
and to use the fact that the saddle point
$\langle \lambda_{\rm s} \rangle$ 
falls 
on an edge of the arcs due to the level repulsion.
Then $\langle \lambda_{\rm s} \rangle$ is
expressed as a function of $g$ only and 
the renormalization group equation takes
a simple linear form.

In $g > g_{*}$ case, assuming a single arc 
configuration at $[-A, A]$ 
and the asymptotics 
$N^{-1}\langle \,{\rm tr}\, (z-\phi)^{-1} \rangle \sim 1/z$
$(|z| \rightarrow \infty)$, we obtain
\begin{equation}
\left\langle \frac{1}{N} \,{\rm tr}\, \frac{1}{z-\phi} \right\rangle 
=\frac{1}{2} \left[
 -z+g z^3 -\left(-1 +\frac{g}{2} A^2 + g z^2 \right)
\sqrt{z^2 - A^2} \right],
\label{1arc}
\end{equation}
where $A^2=\frac{2}{3g}(1+ \sqrt{1+12g})$.
Knowing that
the saddle point is given by the edge of the arc
$\langle \lambda_{\rm s} \rangle = \pm A$,  
the $G$-function (\ref{G}) then takes the form:
\begin{equation}
G\left(g, \frac{dF}{dg} \right) 
= -g \frac{dF}{dg}
-\frac12 -\frac{A^2}{8}-\log \frac{A^2}{4} .
\label{G1}
\end{equation} 
In the leading order of $1/N$, 
the renormalization group equation (\ref{nrge}) becomes
\begin{equation}
2 F(g) = G \left( g, \frac{dF}{dg} \right).
\label{LRGE}
\end{equation}
The solution to this differential equation 
with the $G$-function (\ref{G1}) is
\begin{equation}
F(g)= -\frac{3}{8} -\frac{5 A^2}{48}-\frac{A^4}{384} 
-\frac{1}{2} \log\frac{A^2}{4} + \frac{C_1}{g^2},
\label{sol1}
\end{equation}
where $C_1$ is an integration constant.

In $g < g_{*}$ case, 
we assume 
the double arc 
configuration at $[-A_2, -A_1] \cup [A_1, A_2]$,
\begin{equation}
\left\langle \frac{1}{N} \,{\rm tr}\, \frac{1}{z-\phi} \right\rangle 
= \frac{1}{2} 
\left(-z+ g z^3 - g z \sqrt{(z^2-A_1 ^2)(z^2 -A_2 ^2)} \right),
\label{2arc}
\end{equation}
where ${A_1}^2=(1- 2\sqrt{g})/g$ and ${A_2}^2 = (1+2\sqrt{g})/g$.
By requiring the continuity from the $g > g_{*}$ case,
we are obliged to choose the outer edge $\pm A_2$ 
as the relevant saddle point
$\langle \lambda_{\rm s} \rangle$.
Then the $G$-function (\ref{G}) takes the form:
\begin{equation}
G\left(g, \frac{dF}{dg} \right) =
 -g\frac{dF}{dg} 
-\frac{1}{2}
+ \frac{1}{2} \log g 
-\frac{1}{4g}
.
\label{G2}
\end{equation}
The solution to the renormalization group equation (\ref{LRGE})
with this $G$-function becomes
\begin{equation}
F(g) = 
-\frac{3}{8} 
+ \frac{1}{4} \log g 
-\frac{1}{4g} 
+ \frac{C_2}{g^2}
\label{sol2}
\end{equation}
with an integration constant $C_2$.
The integration constants $C_1$ and $C_2$ 
must be the same
because of the continuity of 
$F(g)$ at $g=g_{*}$.
On the other hand, $C_2$ must vanish 
since 
the free energy should be dominated by
the minima of the potential
in the small-$g$ limit,
\begin{equation}
F(g) \sim V\left(\pm \frac{1}{\sqrt{g}}\right)=-\frac{1}{4g}
\ \ \ \ \ (g\searrow 0).
\end{equation}
This result (\ref{sol1}) and (\ref{sol2})
with $C_1=C_2=0$ is identical to that obtained in 
ref.\cite{Shimamune:phasesep}\footnote{Its free energies err:
$E_{\rm B}$ misses a constant $\frac12 \log 2$;
$E_{\rm C}$ should read (in our convention (\ref{shimamune}))
$
-{3\over 8} + {{-1 + {\sqrt{1 - 15\,g}}}\over {675\,{g^2}}} + 
  {{-33 - 10\,{\sqrt{1 - 15\,g}}}\over {180\,g}} - \frac12
 \, {{\log ({{1 - {\sqrt{1 - 15\,g}}}\over {15\,g}})}}.
$
}
using BIPZ method.

\subsection{Inverse-gaussian fixed point}
The concept
of flow is indispensable in the renormalization group analysis.
However, the nonlinearity of
our exact renormalization group equation (\ref{nrge}) 
with respect to
${\partial F}/{\partial g}$ makes an idea of flow ambiguous.
Thus we wish to define an approximate flow which visualizes 
the phase structure correctly. 
Indeed there is such a sensible
definition of flow.
\nopagebreak[1]

We define an approximate $\beta$-function
expanded around $\bar{g}$ as
\begin{equation}
\beta (g) 
:= \left.\frac{\partial G(g,a)}{\partial a} 
\right|_{a=\bar{a}_1+2\bar{a}_2(g-\bar{g})}
\label{linearbeta}
\end{equation}
where $\bar{a}_1 = 
\frac{dF}{dg} ( \bar{g})$,
$\bar{a}_2 = \frac12 
\frac{d^2 F}{d g^2}
( \bar{g})$,
and shall check the validity of this definition.
Let us first note that 
\begin{equation}
  \beta^{\rm exact}(g) 
:= \left.\frac{\partial G(g,a)}{\partial a} 
\right|_{a= \frac{dF}{dg} (g)}
\label{exactbeta}
\end{equation}
is the `exact' beta function in the sense that the free energy $F$ obeys
a linear equation
\begin{equation}
  \left[ N \frac{\partial}{\partial N} 
-  \beta^{\rm exact}(g)  
\frac{\partial}{\partial g} + 2 \right] F(N,g) = r(g).
\end{equation}
The above `exact' beta function is practically useless
because it requires the full information of the free energy.
It is desired to extract correct qualitative information 
(the phase structure of the model or the universality classification of
phase transitions) by the use
of the renormalization group equation (\ref{nrge}) in case
the exact form of the free energy is not available.
On the other hand, the approximated $\beta$-function (\ref{linearbeta})
is obtained by replacing 
$d F/d g$
with its truncated expansion around $\bar{g}$. 
We can determine the coefficients $a_1$ and $a_2$ by solving
(10) 
even if full information of $F$ is not available. 
Thus we can calculate $\beta
(g)$ for arbitrary $g$ only if
eqs.(10) are solved at a single
point $\bar{g}$.

Since we have kept the linear term in $g$ in 
the power series expansion of 
$d F/d g$,
the zeroth and the first derivatives of $G$ at 
$g=\bar{g}$ are exactly reproduced.
Therefore $\beta (g) $ enjoys the
following property\footnote{Previously we have proposed 
in \cite{HiItNiSa:oneandtwo} a linearized $\beta$-function
\[
  \beta^{\rm linear}(g) 
:= \left.\frac{\partial G}{\partial a} \right|_{a=\bar{a}_1}.
\]
Though this beta function vanishes at a fixed point $\bar{g}$,
its first derivative does not give the exact value of 
the critical exponent
because of the missing second term in
(\ref{exponentcond}).}:
{\em
If $\bar{g}$ is a fixed point of the flow 
{\em (}i.e.~solves eqs.{\em (10))},
then}
\setcounter{sub}{1}
\renewcommand{\theequation}{\arabic{equation}\alph{sub}}
\begin{eqnarray}
  \beta
(\bar{g}) & = & 0\ , \\
\addtocounter{equation}{-1}
\addtocounter{sub}{1}
  \frac{\partial \beta
}{\partial g}(\bar{g})
  &  = &   \frac{2}{\gamma}\ .
  \end{eqnarray}
Even if $\bar{g}$ does not solve the equations (10),
the definition (\ref{linearbeta}) makes sense and $\beta (g)$ 
approximates the character of the renormalization group 
transformation in the vicinity of $\bar{g}$. 
However, expansion around such a $\bar{g}$ is not very useful 
because of the lack of algorithm to determine $a_1(\bar{g})$.
\renewcommand{\theequation}{\arabic{equation}}

The definition (\ref{linearbeta}) is 
easily extended to multi-coupling case as:
\begin{equation}
  \beta_k (g_j) 
:= \left.\frac{\partial G(g_j,a_j)}{\partial a_k} 
\right|_{a_j=\bar{a}_j
           +2\sum_\ell \bar{a}_{j\ell}(g_\ell-\bar{g}_\ell)}
\label{linearbeta-multi}
\end{equation}
where $\bar{a}_j = a_j( \bar{g}_k)$. Notations $a_j$
and $a_{j\ell}$ are those used in ref.\cite{HiItNiSa:oneandtwo}.

\begin{figure}[htbp]
  \begin{center}
    \leavevmode
    \epsfxsize=200pt
    \epsfbox{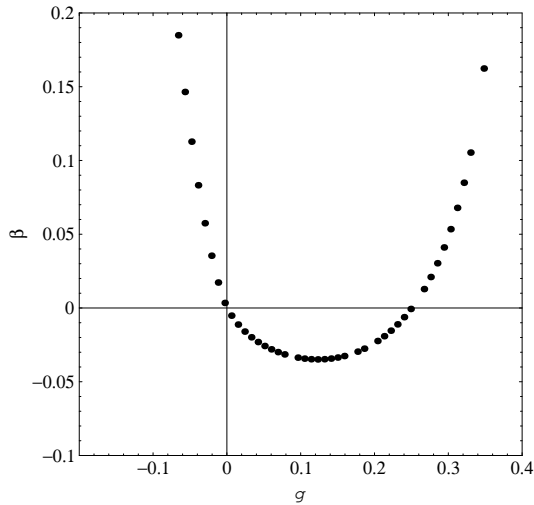}

    \caption{Approximate $\beta$-function 
             (expanded around $g_*=1/4$).}
    \label{linearizedbetashim}
  \end{center}
\end{figure}

In Fig.\ref{linearizedbetashim}, 
the $\beta$-function
for the model (\ref{shimamune}) expanded around the 
merging fixed point 
$g_*=1/4$ is plotted. 
We observe that the approximate $\beta$-function
is smooth everywhere even though the matrix integral is
ill-defined for $g\leq 0$.
It vanishes at the very vicinity of
the origin with a negative exponent.
This indicates that the origin is an attractive fixed point.
The level density becomes two infinitely separated semi-circles
when approaching this point.

It is argued by Br\'ezin and Zee \cite{BrZe:universal}, 
although not rigorously proven,
that the 
universality of the large-$N$ connected two-level correlator
be attributed to the presence of attractive
gaussian fixed point of the renormalization group 
transformation.
If we accept this argument, then 
the attractive inverse-gaussian fixed point in the model
(\ref{shimamune}) plays the same role
for the new universality 
for the multiple-arc matrix model recently found in
ref.\cite{AmAk:newuniversal}.

\subsection{Stability of the gaussian fixed point}
To strengthen our assertion that each fixed point corresponds to a
universality class of large-$N$
two-level correlators,
we present here the analysis of stability of 
the gaussian potential
in the model\footnote{
A renormalization group analysis of the stability of the gaussian
potential has been done by
Y.~Morita et.~al.~({\em Phys.~Rev.} {\bf B52} (1995) 4716)
based on mapping from matrix models to vector models. 
They have imposed an arbitrary restriction  
to a smaller subspace of coupling constants 
which is inconsistent with the renormalization group flow,
yielding incorrect values of the exponents.
}
\begin{equation}
V(\phi)  =  \sum_{k=2}^\infty \frac{g_k}{k} \phi^k,\ \ \ 
g_2= + 1.
\label{multi-coupling}
\end{equation}
We evaluate the derivatives of the $G$-function (\ref{g-function}) at 
the gaussian point $g_k=0$ $(k=3, 4, \cdots)$.
We immediately find that 
\[
\left.\frac{\partial}{\partial a_\ell}
\langle\lambda_{\rm s}\rangle\right|_{g_j=0}
=
\left.\frac{\partial}{\partial a_\ell}
\left\langle\frac1N\,{\rm tr}\,
\frac{1}{z-\phi}\right\rangle\right|_{g_j=0}=0  
\]
implying
\begin{equation}
  G_{,a_i}(g_j=0,a_n) =   G_{,a_\ell,a_k}(g_j=0,a_n) =  0 
\end{equation}
for arbitrary $a_n$.
Thus the gaussian point is always a fixed point.

To obtain the exponents on the gaussian fixed point, 
we need to
calculate $G_{,g_k,a_\ell}$. 
The second term in
(\ref{g-function}) contributes $(1-k/2) \delta_{\ell k}$
to it.
It can also be shown that 
\[
\left.\frac{\partial^2}{\partial g_k \partial a_\ell}
\left\langle \frac1N\,{\rm tr}\,\frac{1}{z-\phi}
\right\rangle\right|_{g_j=0}=0  
\ \ \mbox{ if }\ \  \ell +2 > k.
\]
Therefore the scaling exponent matrix 
$
G_{\ell k} := G_{,g_k,a_\ell} + 2 a_{kr} G_{,a_\ell,a_r}
$
is upper triangular with eigenvalues  
\begin{equation}
  G_{kk}( 0,0,\ldots)  =  1-\frac{k}{2}
\ \ \ 
(k =3, 4, \cdots).
\end{equation}
The negativity of the exponents
indicates the stability of the gaussian fixed point 
in the entire coupling constant space (\ref{multi-coupling})
in accord with the universality of the 
large-$N$ two-level correlator.

In the $g_2 = -1$ case, 
the model at the inverse-gaussian point is ill-defined
and therefore the general algorithm cannot extract 
any critical exponents around this point.
Nevertheless, 
the model is well-defined except at the inverse-gaussian point.
The inverse-gaussian point is a fixed point 
in the sense of the limiting procedure
\begin{equation}
\lim_{g \rightarrow 0} G_{,a}(g,a) = 0.
\end{equation} 
The stability of this fixed point is
guaranteed by the negative
definiteness of the extended $\beta$-function $G_{,a}$
for $0<g<\frac{1}{4}$.     

\subsection{Summary}
This letter is
devoted mainly to the application of
the method of large-$N$ renormalization group
to a matrix model with a double-well potential.
The advantage of our method is that
the arc configuration need not be specified from the outset, 
but rather a single
renormalization group equation (\ref{G})
(and its approximated $\beta$-function) 
encompasses all possible arc configurations.
As a direct consequence of it,
we have extracted
exact values of the critical (fixed) point and 
exponent 
associated with the splitting/merging of the arc(s).
Furthermore, 
by the use of approximate $\beta$-function
we have found
a novel inverse-gaussian fixed point,
whose stability may guarantee the large-$N$ universality of 
double-arc matrix ensemble. 
We have also calculated the scaling exponents
around the gaussian fixed point for 
a matrix model with a generic potential.
Their negativity, i.e.~the stability of the gaussian point
against perturbation of potential preserving
the single-arc configuration,
supports the large-$N$ universality of (usual single arc)
matrix ensembles.

We wish to utilize our results to 
the random matrix analysis of
disordered physical systems.

\renewcommand{\thesubsection}{Acknowledgments}
\subsection{}
S.H.~thanks S.~Hikami for discussions.
S.M.N.~thanks G.~Akemann for explaining his work
and for discussions.
This work is supported in part by 
Grant-in-Aid for Scientific Research 
(S.H., No.~08740196, N.S., No.~05640334) 
and Grant-in-Aid for Scientific Research
for Priority Areas (N.S., No.~05230019) from the Ministry of
Education, Science, and Culture, 
by CREST from Japan Science and Technology Corporation (S.H.),
and by JSPS Postdoctoral Fellowships for Research Abroad (S.M.N.).


\begin{thebibliography}{99}
\bibitem{AmAk:newuniversal}
J.~{Ambj{\o}rn} and G.~Akemann,
\newblock {\em J.~Phys.} {\bf A29} (1996) L555;\\
G.~Akemann,
\newblock {\em Nucl.~Phys.} {\bf B482} (1996) 403.
\bibitem{BrZe:universal}
E.~{Br\'ezin} and A.~Zee,
\newblock {\em Nucl.~Phys.} {\bf B402} (1993) 613.
\bibitem{BrZi:RGMat}
E.~{Br\'{e}zin} and J.~{Zinn-Justin},
\newblock {\em Phys.~Lett.} {\bf B288} (1992) 54.
\bibitem{HiItNiSa:Matrix}
S.~Higuchi, C.~Itoi, S.~Nishigaki, and N.~Sakai,
\newblock {\em Phys.~Lett.} {\bf B318} (1993) 63.
\bibitem{HiItNiSa:oneandtwo}
S.~Higuchi, C.~Itoi, S.~Nishigaki, and N.~Sakai,
\newblock {\em Nucl.~Phys.} {\bf B434} (1995) 
\nolinebreak[1] 283.
\bibitem{David:c>1}
F.~David, Saclay preprint 
SPhT/96-112, hep-th/9610037, v2.
\bibitem{ShVe:RMTsumrule}
E.~V.~Shuryak and J.~J.~M.~Verbaarschot,
\newblock {\em Nucl.~Phys.} {\bf A560} (1993) 306.
\bibitem{JaSeVe:zerovertuality}
A.~D.~Jackson, M.~K.~\c{S}ener, and J.~J.~M.~Verbaarschot,
\newblock {\em Nucl.~Phys.} {\bf B479} (1996) 707.
\bibitem{Shimamune:phasesep}
Y.~Shimamune,
\newblock {\em Phys.~Lett.} {\bf B108} (1982) 407.
\bibitem{BrItPaZu:planar}
E.~Br{\'{e}}zin, C.~Itzykson, G.~Parisi, and J.-B. Zuber,
\newblock {\em Commun.~Math.~Phys.} {\bf 59} (1978) 35.
\end{thebibliography}
\end{document}